\documentclass[twocolumn,aps,prc,floatfix]{revtex4}
\usepackage{graphicx}
\usepackage{dcolumn}
\usepackage{bm}
\usepackage[version=4]{mhchem}
\usepackage{color}
\def\es0{$E_{\rm sym}(\rho_0)$}

\def\us0{$U_{\rm sym}(\rho_0,k_F)$~}

\def\l0{$L(\rho_0)$~}

\begin{document}

\title{Properties of First-Order Hadron-Quark Phase Transition from Inverting Neutron Star Observables}

\author{Nai-Bo Zhang$^{1}$\footnote{naibozhang@seu.edu.cn} and Bao-An Li$^2$\footnote{Bao-An.Li@Tamuc.edu}}

\affiliation{$^1$School of Physics, Southeast University, Nanjing 211189, China}
\affiliation{$^2$Department of Physics and Astronomy, Texas A$\&$M University-Commerce, Commerce, TX 75429-3011, USA}
\date{\today}

\setcounter{MaxMatrixCols}{10}

\begin{abstract}

By inverting the observational data of several neutron star observables in the three dimensional parameter space of the constant speed of sound (CSS) model while fixing all hadronic Equation of State parameters at their currently known most probable values, we constrain the three parameters of the CSS model and their correlations. Using two lower radius limits of $R_{2.01}=11.41$ km and $R_{2.01}=12.2$ km for PSR J0740+6620 obtained from two independent analyses using different approaches by the Neutron Star Interior Composition Explorer (NICER) Collaboration, the speed of sound squared $c_{\rm QM}^2$ in quark matter is found to have a lower limit of $0.35$ and $0.43$ in unit of $c^2$, respectively, above its conformal limit of $c_{\rm QM}^2<1/3$. An approximately linear correlation between the first-order hadron-quark transition density $\rho_t$ and its strength $\Delta\varepsilon$ is found. Moreover, the presence of twin star is deemed improbable by the present work.
\end{abstract}
\maketitle


\section{Introduction}

Novel phenomena are expected to occur in dense neutron-rich matter existing naturally in neutron stars. However, their signatures remain inconclusive and have attracted much effort in both nuclear astrophysics and astronomy. In particular, properties of possible phase transitions in dense matter can affect the kilonova explosions and gamma-ray bursts associated with binary neutron star mergers. Their signatures may be identified from the post merger high-frequency gravitational wave signals of high-mass binaries using the next generation gravitational wave detectors \cite{Orsaria19,Weih20}, such as the Cosmic Explorer \cite{Reitze19}, Einstein Telescope \cite{Punturo10}, or Neutron Star Extreme Matter Observatory \cite{Ackley20}.

The hadron-quark phase transition is expected to occur at high baryon densities, leading to the formation of hybrid stars. However, there is still no consensus on properties of such phase transition, such as its onset density, nature (first-order or smooth crossover), and strength (jump in energy density). For example, some work found that the phase transition may occur around 3$\sim$4 times the saturation density of nuclear matter ($\rho_0=0.16$ fm$^{-3}$) \cite{Tang21,Yong22} while a significantly lower transition density is favoured in Refs. \cite{Miao20,Liang21,Xie21,Somasundaram22}. Clearly, further research is needed to clarify many remaining issues and improve our understandings about the phase transition. Fortunately, recent observations of neutron stars facilitate efforts in this direction and enable the community to move further close to realizing the ultimate goal of determining the nature and Equation of State (EOS) of dense neutron-rich matter. In particular, the maximum observed mass of neutron stars has increased from $2.08\pm0.07$ M$_\odot$ \cite{Mmax,Fonseca21} to $2.35\pm0.17$ M$_\odot$ \cite{Romani22}. Two independent analyses using different approaches by the Neutron Star Interior Composition Explorer (NICER) Collaboration have reported that the  radius of PSR J0740+6620 with a mass of $(2.08\pm 0.07)$M$_{\odot}$ is: $R_{\rm PSR~J0740+6620}=13.7_{-1.5}^{+2.6}$ km \cite{Miller21} and $R_{\rm PSR~J0740+6620}=12.39_{-0.98}^{+1.30}$ km \cite{Riley21}, respectively. Additionally, the LIGO and Virgo Collaborations have found that the tidal deformability of canonical neutron stars is about $70<\Lambda_{1.4}<580$ at 90\% confident level \cite{LIGO18}. These observations have provided the much needed data, albeit still very limited and some have large uncertainties, for better understanding properties of neutron star matter. Indeed, they have been used repeatedly in various analyses in the recent literature and very interesting physics has been extracted.

The perturbative quantum chromodynamics (pQCD) can describe quark matter accurately when the baryon density is larger than about 40$\rho_0$. Very interestingly, it was predicted that the speed of sound squared of quark matter has the so-called conformal limit of $c^2_{\rm QM}(\rho)<1/3$ \cite{Kurkela10,Gorda18}. However, how the $c^2_{\rm QM}(\rho)$ varies with density, whether and how the conformal limit may be reached in neutron stars is still an open question. For example, some studies (see, e.g., Refs. \cite{Kurkela14,Bedaque15,Alsing18,Godzieba21}) have shown that satisfying the conformal limit at any density is contradictory to the observations of massive neutron stars \cite{Mmax,Fonseca21}, and thus a bump in the variation of $c^2_{\mathrm{QM}}(\rho)$ with increasing density is expected. Generally speaking, current predictions or assumptions about $c^2_{\mathrm{QM}}(\rho)$ are rather model-dependent \cite{Tan20,Ecker22}. One extreme assumption is that the $c^2_{\mathrm{QM}}(\rho)$ is a constant. Such assumption is used in the constant speed of sound (CSS) model \cite{Alford13,Chamel13,Zdunik13} in describing a first-order hadron-quark phase transition in hybrid stars. The CSS model coupled with various hadronic EOSs has been used in studying properties of hybrid stars extensively \cite{Alford13,Ayriyan15,Alford15,Chatziioannou20,Han20,Miao20,Xie21,LiJJ21,Li22,Drischler22}. It has three parameters: the transition pressure $p_t$ (or transition density $\rho_t$), discontinuity in energy density $\Delta \varepsilon$, and the $c_{\rm QM}$ assumed to be density-independent. In fact, the latter assumption is consistent with predictions of some Nambu-Jona-Lasinio models \cite{Agrawal10,Bonanno12,Lastowiecki12,Zdunik13}, perturbation theories \cite{Kurkela10,Kurkela10b}, or bag-model-like EOSs \cite{Traversi20,Traversi21}. Thus, simultaneously constraining the above three parameters using the available observational data of neutron stars can enhance our understanding of quark matter in hybrid stars.

In this work, by inverting the observables of neutron stars in the CSS model's 3-dimensional (3D) parameter space while fixing all hadron matter EOS parameters at their currently known most probable values, we investigate how/if the available neutron star observational data can constrain the CSS model parameters and their correlations. The rest of this paper is organized as follows: The model EOSs for hadronic and quark matter are described in section \ref{sec2}. The results of this work are discussed in section \ref{sec3}. Our conclusions are summarized in section \ref{sec4}.

\section{Meta model Equation of states for hybrid stars}\label{sec2}
In the present work, an EOS of hadronic matter consisting of nucleons, electrons, and muons ($npe\mu$) at $\beta$-equilibrium is connected to the CSS model.
While all parameters in the hadronic part of the EOS are fixed at their currently known most probable values, all three parameters in the CSS model are considered essentially as free. We then invert the neutron star observables in the 3D parameter space of CSS. In this sense, the CSS model is used as a meta model. For completeness and ease of following discussions, here we recall briefly the main features of the hadronic EOS we use for the $npe\mu$ matter at $\beta$-equilibrium and the CSS model for a first-order hadron-quark phase transition and the EOS of quark matter in hybrid stars.

\subsection{An EOS for hadronic matter in neutron stars}
The EOS of $npe\mu$ matter in neutron stars at $\beta$-equilibrium can be constructed (see, e.g. \cite{Zhang18}) using the parameterized energy per nucleon $E_{0}(\rho)$ of symmetric nuclear matter (SNM) and the nuclear symmetry energy $E_{\rm sym}(\rho)$:
\begin{equation}\label{E0-taylor}
  E_{0}(\rho)=E_0(\rho_0)+\frac{K_0}{2}(\frac{\rho-\rho_0}{3\rho_0})^2+\frac{J_0}{6}(\frac{\rho-\rho_0}{3\rho_0})^3,
\end{equation}
\begin{eqnarray}\label{Esym-taylor}
    E_{\rm{sym}}(\rho)&=&E_{\rm{sym}}(\rho_0)+L(\frac{\rho-\rho_0}{3\rho_0})\nonumber\\
    &+&\frac{K_{\rm{sym}}}{2}(\frac{\rho-\rho_0}{3\rho_0})^2
  +\frac{J_{\rm{sym}}}{6}(\frac{\rho-\rho_0}{3\rho_0})^3.\nonumber\\
\end{eqnarray}
Around the saturation density $\rho_0$ of SNM, the parameters in the above equations have the same meaning as the Taylor expansion coefficients of nuclear energy density functionals. They are widely used in studying properties of neutron stars and nuclei as well as their mergers and collisions. Moreover, the Eqs. (\ref{E0-taylor}) and (\ref{Esym-taylor}) can be seen simply as parameterizations and the coefficients are free parameters especially at high densities when the Taylor expansions do not converge. Much efforts have been made by the community to constrain the ranges of these parameters within various approaches (including both forward modelings and backward inferences) using data from terrestrial experiments,
astrophysical observations, and theoretical predications. Examples of using this hadronic EOS in studying several properties of neutron stars can be found in our previous publications \cite{Zhang18,Zhang19,Zhang19a,Zhang19b,Zhang2020,Zhang2021,Zhang22,Xie19,Xie20,Xie21}.

Based on many terrestrial experiments and astrophysical observations as well as theoretical calculations available before 2016, the binding energy $E_0(\rho_0)$ and incompressibility $K_0$ at $\rho_0$ have been constrained to $E_0(\rho_0)=-15.9\pm0.4$ MeV and $K_0=240\pm20$ MeV \cite{Garg18,Shlomo06}, while the symmetry energy $E_{\rm sym}(\rho_0)$ and its slope $L$ at $\rho_0$ are constrained to $E_{\rm sym}(\rho_0)=31.7\pm3.2$ MeV and $L=58.7\pm28.1$ MeV \cite{Li13,Oertel17}, respectively. Recently, Ref. \cite{Li21} surveyed 24 new analyses of neutron stars between GW179817 and 2021. It was found there that the available analyses gave an average value of $L=57.7\pm19$ MeV and that of the curvature of symmetry energy $K_{\rm sym}=-107\pm88$ MeV at 68\% confidence level, respectively. The later is consistent with $K_{\rm sym}=-100\pm100$ constrained in Refs. \cite{Mondal17,Margueron18,Somasundaram21,Grams22}. In addition, within the same framework of the present work, $K_{\rm sym}=-230^{+90}_{-50}$ MeV was found in Bayesian analyses of the available neutron star observables \cite{Xie19}.

As a parameter characterizing the stiffness of SNM EOS at densities above about (2-3)$\rho_0$, $J_0$ is constrained to $J_0=-190\pm40$ MeV at 68\% confidence level based on Bayesian analyses of neutron star radii from LIGO and NICER observations \cite{Xie19,Xie20}, $J_0=-180^{+100}_{-110}$ MeV from a Bayesian analysis of nuclear collective flow in relativistic heavy-ion collisions \cite{Xie21}, $-180<J_0<200$ MeV by combining the observed mass of PSR PSR J0740+6620 and causality condition \cite{Zhang19b}. However, very few constraint on the $J_{\rm sym}$ characterizing the stiffness of symmetry energy at densities above about (2-3)$\rho_0$ has been obtained so far \cite{Somasundaram21}.

Based on the information provided above, in the present work, we use $E_0(\rho_0)=-15.9$ MeV, $E_{\rm sym}(\rho_0)=31.7$ MeV, $K_0=240$ MeV, $L=58.7$ MeV, $K_{\rm sym}=-230$ MeV, $J_0=-190$, and $J_{\rm sym}=300$ MeV.
To our best knowledge, currently they are approximately the most probable values of these parameters. Once the parameters in Eqs. (\ref{E0-taylor}) and (\ref{Esym-taylor}) are given, a unique EOS for $npe\mu$ matter in neutron stars at $\beta$-equilibrium can be obtained from the energy density
\begin{equation}\label{lepton-density}
  \varepsilon(\rho, \delta)=\rho [E(\rho,\delta)+M_N]+\varepsilon_l(\rho, \delta),
\end{equation}
where $M_N$ represents the average nucleon mass, $E(\rho,\delta)=E_0(\rho)+E_{\rm{sym}}(\rho)\cdot\delta ^{2}+\mathcal{O}(\delta^4)$ is the average energy per nucleon of neutron-rich nuclear matter with isospin asymmetry  $\delta=(\rho_n+\rho_p)/\rho$, and $\varepsilon_l(\rho, \delta)$ denotes the lepton energy density \cite{Oppenheimer39}. The particle densities (consequently the density profile of isospin asymmetry $\delta(\rho)$) can be obtained by solving the $\beta$-equilibrium condition $\mu_n-\mu_p=\mu_e=\mu_\mu\approx4\delta E_{\rm{sym}}(\rho)$ where $\mu_i=\partial\varepsilon(\rho,\delta)/\partial\rho_i$ and charge neutrality condition $\rho_p=\rho_e+\rho_\mu$. Then the pressure becomes barotropic and can be calculated from:
\begin{equation}\label{pressure}
  P(\rho)=\rho^2\frac{d\varepsilon(\rho,\delta(\rho))/\rho}{d\rho}.
\end{equation}
Similarly, the  energy density $\varepsilon(\rho, \delta(\rho))\rightarrow \varepsilon(\rho)$ becomes barotropic and the resulting EOS $P(\varepsilon)$ is used in solving the Tolman-Oppenheimer-Volkoff (TOV) equation.

To exclude effects of the crust on properties of hybrid stars, we fix the crust-core transition density at $0.08$ fm$^{-3}$ and choose the NV EOS \cite{Negele73} for the inner crust and the BPS EoS \cite{Baym71b} for the outer crust. This choice is consistent with the hadronic EOS parameters selected above \cite{Zhang19a} and it does not affect any conclusion we make in this work.

\subsection{A meta model for hadron-quark phase transition and quark matter}

With the increase of density, a phase transition from hadronic to quark matter is expected to happen. We adopt the CSS model of Alford, Han and Prakash \cite{Alford13} assuming the hadron-quark phase transition is first order. In this model, the entire EOS of hybrid stars can be described by \cite{Alford13,Chamel13,Zdunik13}:
\begin{equation}
\varepsilon(\rho)= \begin{cases}\varepsilon_{\mathrm{HM}}(\rho) & \rho<\rho_{t} \\ \varepsilon_{\mathrm{HM}}\left(\rho_{t}\right)+\Delta \varepsilon+c_{\mathrm{QM}}^{-2}\left(p-p_{t}\right) & \rho>\rho_{t}\end{cases}
\end{equation}
where $\varepsilon_{\mathrm{HM}}(p)$ is the energy density of hadronic matter below the transition density $\rho_t$. Since we can run through the whole 3D parameter space in $\rho_t-\Delta \varepsilon-c_{\mathrm{QM}}^{-2}$, the CSS model can be considered as a meta model. It can mimic many features of more microscopic quark matter models and it has been widely used in studying hybrid stars as mentioned earlier \cite{Alford13,Ayriyan15,Alford15,Chatziioannou20,Han20,Miao20,Xie21,LiJJ21,Li22,Drischler22}. Once the hadron EOS is given, properties of hybrid stars are solely determined by the three CSS model parameters. 

In previous studies employing the CSS model, people normally consider several representative speed of sound values (e,g., $c^2_{\rm QM}=1/3$ and $c^2_{\rm QM}=1$ in Refs. \cite{Alford15,Miao20}). However, in this work, we hope to establish direct connections between the observables and the CSS model parameters, enabling us to simultaneously constrain all three CSS parameters using the observable data. Consequently, we treat $c^2_{\rm QM}$ as a free parameter that needs to be determined through inverting neutron star observables in the 3D CSS model parameter space, which will be introduced in subsection \ref{2-D}. Surprisingly, as we shall show below, $c^2_{\rm QM}$ is constrained to the range of $0.35<c^2_{\rm QM}<1$, which is consistent with the widely used representative speeds of sound.

To satisfy the causality condition, we require $c^2_{\rm QM}\leq 1$. Also we choose a lower limit for $c^2_{\rm QM}\geq 0.1$ as a hybrid star may not exist if the EOS of quark matter is too soft. As the hybrid branch may not exits for large $\Delta\varepsilon$, we also require that the discontinuity in energy density $\Delta\varepsilon$ is smaller than 500 MeV. As nuclear matter is stable and the phase transition is not expected around the saturation density $\rho_0$, $\rho_t>\rho_0$ is set as a loose lower limit for the transition density while several higher values for this limit have been proposed/found based on various analyses in the literature, e.g., $\rho_t/\rho_0>1.84$ \cite{Tang21}, $\rho_t/\rho_0>1.3\sim1.5$ \cite{Miao20}, and $\rho_t/\rho_0>1.7$ \cite{Christian20}. In short, the ranges of the three CSS parameters are selected as: $\rho_t/\rho_0>1$, $0<\Delta\varepsilon<500$ MeV, and $0.1<c^2_{\rm QM}<1$. As we will see below, the above ranges are large enough for inverting the presently available neutron star observational data without prior biases.

\subsection{The neutron star observational data used}

Much progress has been achieved in the observations of neutron stars in recent years especially since GW170817. For instance, the mass of PSR J0740+6620 has been updated to be $2.08\pm0.07$ M$_\odot$ at 68\% confidence level \cite{Mmax,Fonseca21}. Here we limit ourselves to studying non-rotating neutron stars. Thus, the fastest and heaviest known galactic neutron star with mass $M=2.35\pm0.17$ M$_\odot$ and frequency $f=709$ Hz \cite{Romani22}, Pulsar PSR J0952-0607 is not included in the present work because it is known that fast rotations can increase appreciably the maximum mass that a given EOS can support. Instead, we choose $M_{\rm max}=2.01$ M$_\odot$ as the {\it minimum} maximum mass (lower limit on the maximum mass on any mass-radius sequence predicted by any EOS) and thus the peaks of the mass-radius curves for all EOSs have to be larger than 2.01 M$_\odot$.

Additionally, considering the two lower limits mentioned earlier for the radius at 68\% confidence level for PSR J0740+6620 from NICER, we choose $R_{2.01}=11.41$ km and $R_{2.01}=12.2$ km as two independent observations of the lower radius limits of neutron stars with a mass of 2.01 M$_\odot$. We purposely exclude the upper limits of $R_{2.01}$ from the two NICER analyses mentioned earlier as they provide less strict constraints on the EOS compared to the observation of the upper limit of $\Lambda_{1.4}$ \cite{Zhang2021}. We thus also use the upper limit of tidal deformability for canonical neutron stars $\Lambda_{1.4}=580$ at 90\% confidence level from GW170817 \cite{LIGO18}. Note that it corresponds to $\Lambda_{1.4}=427$ at 68\% confidence level (which is used here to unify the confidence level for all the observables considered). Similar to the upper limit of $R_{2.01}$, the lower limit of $\Lambda_{1.4}$ from GW170817 is not as restrictive as the lower limit of $R_{2.01}$ in constraining the EOS, and it is thus not considered in the following analyses. Therefore, all EOSs selected in this work should satisfy: $M_{\rm max}>2.01$ M$_\odot$, $R_{2.01}>11.41$ km or $R_{2.01}>12.2$ km, and $\Lambda_{1.4}<427$.

\begin{figure*}[ht]
  \centering
   \resizebox{0.33\textwidth}{!}{
  \includegraphics[bb=10 10 630 630]{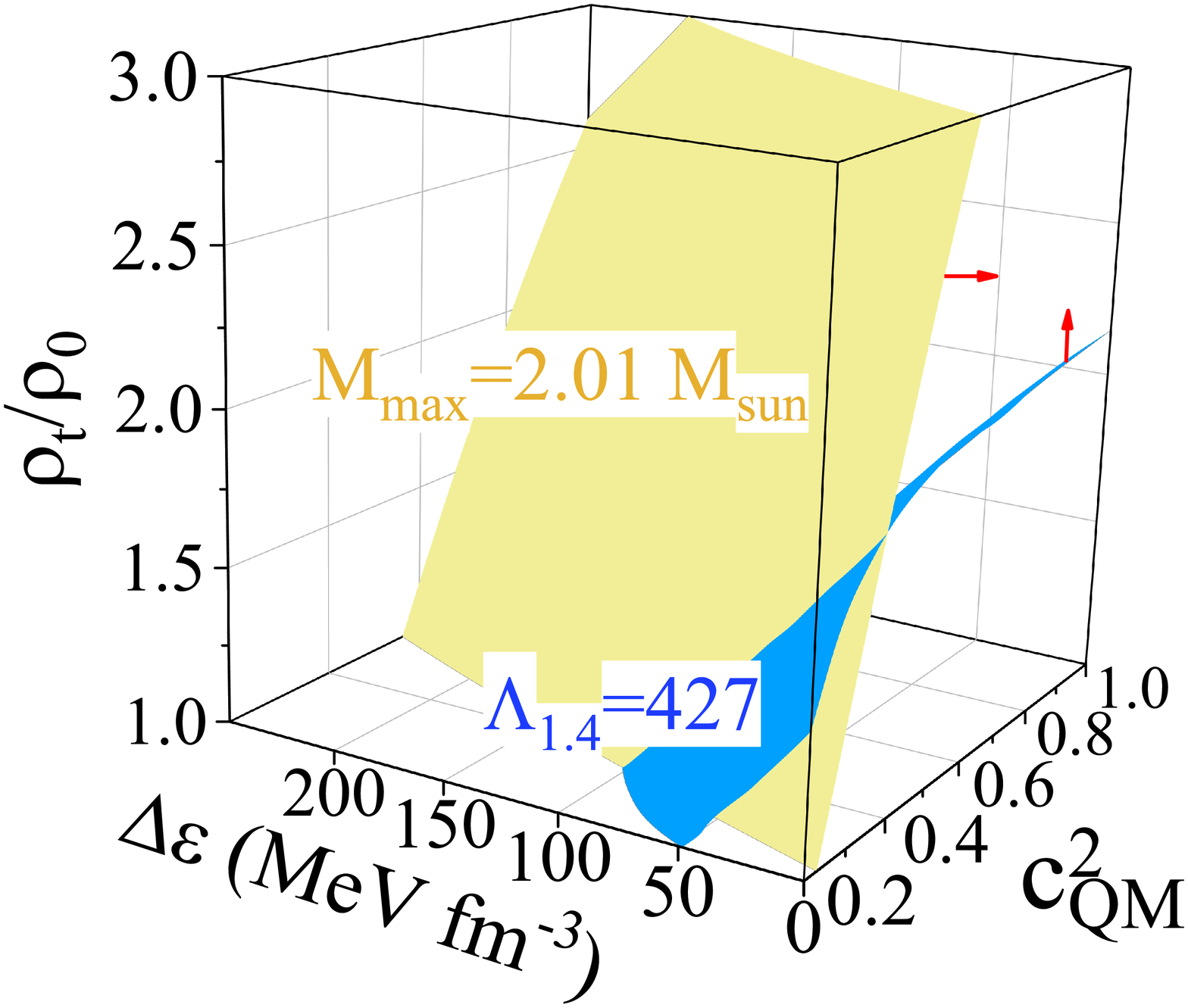}
}
\hspace{2cm}
 \resizebox{0.33\textwidth}{!}{
  \includegraphics[bb=10 10 630 630]{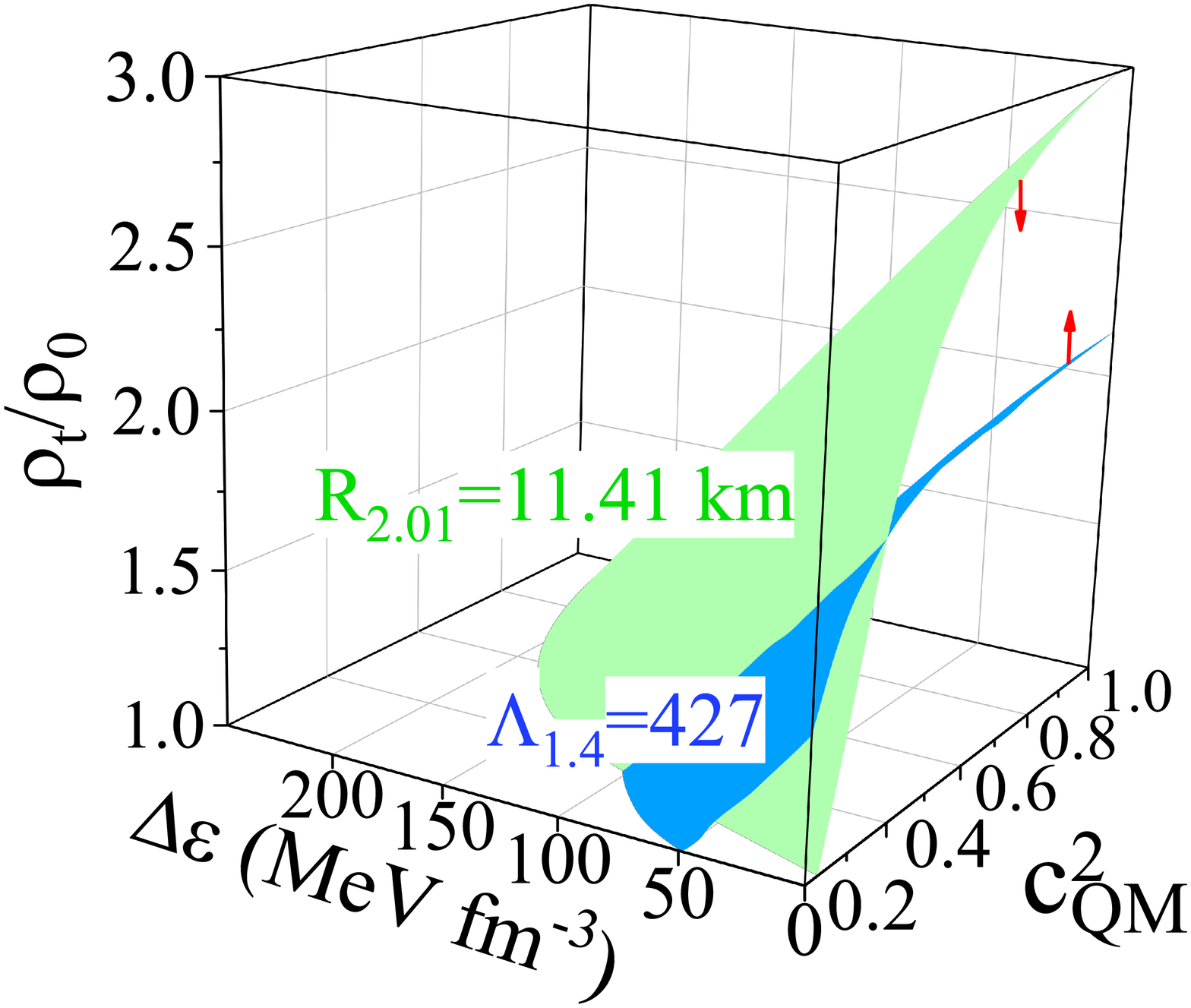}
}
 \resizebox{0.33\textwidth}{!}{
  \includegraphics[bb=10 10 630 630]{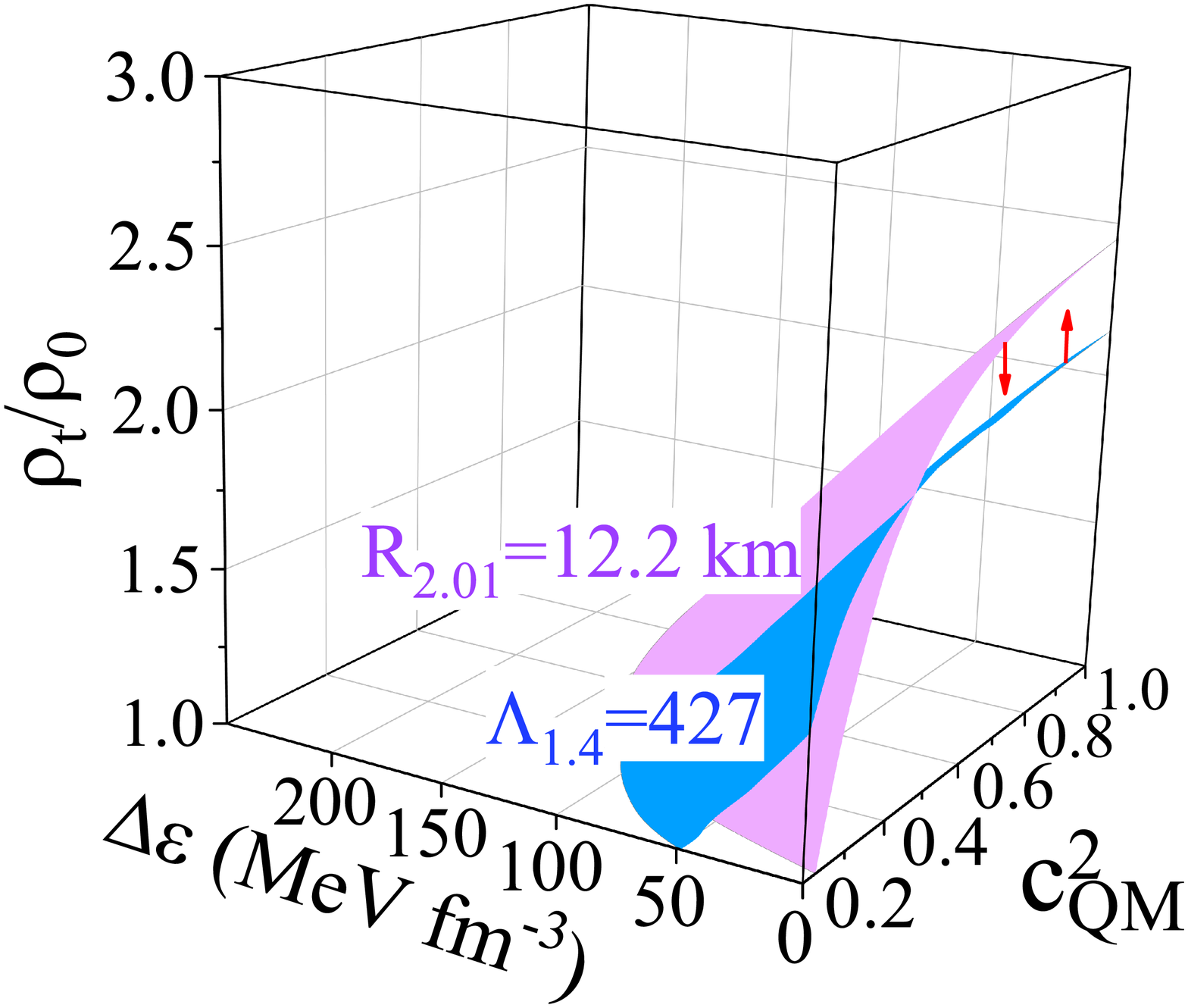}
  }
  \caption{The combined constraints of mass, radius, and tidal deformbility in the 3D parameter space of $\rho_t/\rho_0-\Delta\varepsilon-c^2_{\rm QM}$. The red arrows show the directions that satisfying the corresponding observations. }\label{Constraint}
\end{figure*}

\subsection{Inversion of neutron star observables in the CSS model's 3D parameter space by brute force}
\label{2-D}

Solving the inverse structure problem of neutron stars, i.e., inferring the internal structure/composition/EOS from neutron star observables, has been a longstanding goal of nuclear astrophysics. While Bayesian statistical inference has been very fruitful in most cases, when the observational data have large uncertainties especially when different kinds of data are combined, it is useful to know what each specific value of an observable may constrain the relevant model parameter space. Given the data discussed in the previous subsection, we use here a brute force approach to directly invert the neutron star observational data in the CSS model's 3D parameter space of $\rho_t/\rho_0-\Delta\varepsilon-c^2_{\rm QM}$. An advantage of this approach is that one can actually visualize effects of each parameter on the observational data or conversely how the latter can limit the parameter space. However, it is only applicable to models with fewer than four parameters.

Technically, instead of generating normally multi-millions of Markov Chain Monte Carlo (MCMC) steps in typical Bayesian analyses, we search the entire parameter space by brute force through three loops of the model parameters. While the posterior probability distribution functions of model parameters in Bayesian analyses describe how statistically the corresponding EOSs can reproduce the observational data normally according to a Gaussian likelihood function,  a given observational data is reproduced within a specified inversion precision (e.g., the precision for $R_{2.01}$ used here is $\pm0.001$ km) by the EOSs found in the inversion. For example, if we want to obtain the parameter sets giving $R_{2.01}=11.41 \pm 0.001$ km, for each set of $\Delta\varepsilon$ and $c^2_{\rm QM}$ values selected (inside the loops of these two parameters), we vary $\rho_t/\rho_0$ in steps of 0.001 (i.e., loop through this parameter range) to find a specific $\rho_t/\rho_0$ value leading to an EOS that gives $R_{2.01}=11.41 \pm 0.001$ km from calling the TOV solver in the loop. Then we loop through the uncertainty ranges of $\Delta\varepsilon$ and $c^2_{\rm QM}$ to find all $\rho_t/\rho_0$ values that yield $R_{2.01}=11.41 \pm 0.001$ km.  After going through all three loops, we can plot a constant observable surface in the 3D parameter space of $\rho_t/\rho_0-\Delta\varepsilon-c^2_{\rm QM}$. Each point on this surface represents a unique EOS. All EOSs on the surface can reproduce the same value of the observable within the specified precision of the inversion process (not that of the data itself).

\section{Results and discussions}\label{sec3}

The combined constraints of mass, radius, and tidal deformability in the 3D parameter space of $\rho_t/\rho_0-\Delta\varepsilon-c^2_{\rm QM}$ are shown in Fig. \ref{Constraint}. The red arrows show the directions satisfying the corresponding observational constraint. We note that the surfaces of $M_{\rm max}=2.01 M_\odot$, $R_{2.01}=11.41$ km, and $R_{2.01}=12.2$ km converge at the front bottom corner if we put them into one plot. This convergence occurs because the radius of a hybrid star with $M_{\rm max}=2.01 M_\odot$ exceeds 12.2 km for soft quark matter EOSs and weak phase transition strengths, and all three constraints are satisfied simultaneously. This convergence disappears gradually as $c^2_{\rm QM}$ and $\Delta\varepsilon$ increase. The aforementioned three surfaces provide the upper and left boundaries for the available parameter space. Additionally, the surfaces of $\Lambda_{1.4}=427$ and $\rho_t/\rho_0=1$ set the lower limit of the available parameter space.  Note here again that we use $\rho_t/\rho_0=1$ as a loose lower limit. The right and back boundaries of the available parameter space are provided by $\Delta\varepsilon=0$ and $c^2_{\rm QM}=1$, respectively. The $\Lambda_{1.4}=427$ surface intersects with other surfaces as $\rho_t/\rho_0$ and $c^2_{\rm QM}$ increase. The enclosed parameter space can satisfy all observational constraints or physical conditions considered. The intersecting lines can be used to constrain the individual parameters or their correlations, which will be discussed below. We can see that while the enclosed space is thin, the parameter uncertainties within it are not strongly constrained by the observational data considered.

Compared to the other three surfaces, the $M_{\rm max}=2.01 M_\odot$ surface is almost vertical and closest to the $\Delta\varepsilon$-$\rho_t$ plane, indicating that $c^2_{\rm QM}$ plays the most important role in determining the $M_{\rm max}$ by controlling the stiffness of quark matter EOS, thereby the maximum mass of hybrid stars. Focusing on the $M_{\rm max}=2.01 M_\odot$ surface, it is seen that when $\Delta\varepsilon$ is less than approximately 230 MeV, $\rho_t/\rho_0$ could exceed 3. However, the simultaneous measurements of mass and radius of PSR~J0740+6620 by NICER have significantly constrained the available parameter space, with the strictest upper limit set by the $R_{2.01}=12.2$ km constraint. Thus, $\rho_t/\rho_0<3$ is used in the following discussions. In addition, the slopes of the surfaces with $R_{2.01}=11.41$ km and $R_{2.01}=12.2$ km decrease with increasing $c^2_{\rm QM}$ apparently. This indicates that the effects of $c^2_{\rm QM}$ on the radii of massive neutron stars decrease for larger $c^2_{\rm QM}$ values. This implies that the radius of a massive neutron star is mainly controlled by $\rho_t/\rho_0$ and $\Delta\varepsilon$ when the EOS of quark matter is already stiff.

\begin{figure}[ht]
  \centering
   \resizebox{0.45\textwidth}{!}{
  \includegraphics{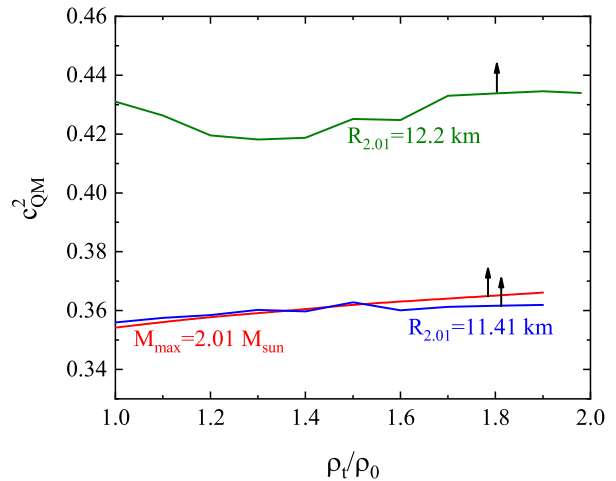}
  }
  \caption{Projections to the $\rho_t/\rho_0-c^2_{\rm QM}$ plane of the intersecting lines between the surface of $\Lambda_{1.4}=427$ and surfaces of $M_{\rm max}=2.01 M_\odot$, $R_{2.01}=11.41$ km, and $R_{2.01}=12.2$ km, respectively. The arrows indicate the directions satisfying the specified observations.}\label{rc}
\end{figure}

To extract quantitatively constraints on the available parameter space and the correlations among the parameters, we now examine the intersecting lines between the surface of $\Lambda_{1.4}=427$ and the surfaces of $M_{\rm max}=2.01 M_\odot$, $R_{2.01}=11.41$ km, and $R_{2.01}=12.2$ km, respectively, and then project them first to the $\rho_t/\rho_0-c^2_{\rm QM}$ plane in Fig. \ref{rc}. These intersecting lines set the lower limits on the $c^2_{\rm QM}$ and the arrows show the directions satisfying the indicated observations. We can see that the lower limits of $c^2_{\rm QM}$ are almost independent of $\rho_t$. The $R_{2.01}=11.41$ km constraint provides almost the same lower limit as $M_{\rm max}=2.01 M_\odot$ because the two surfaces still merge together around $c^2_{\rm QM}=0.35$, if we combine the three boxes in Fig.~\ref{Constraint} together. This is because the radius of a hybrid star with $M_{\rm max}=2.01 M_\odot$ is always larger than 11.41 km for small values of $\Delta\varepsilon$ and $c^2_{\rm QM}$. With the tighter constraint of $R_{2.01}=12.2$ km, the lower limit of $c^2_{\rm QM}$ increases from about 0.35 to 0.43 apparently. This indicates that measuring accurately the radii of massive neutron stars can help constrain tightly the lower limit of $c^2_{\rm QM}$. The large value of $c^2_{\mathrm{QM}}$ is consistent with the findings of some other analyses of the maximum mass of neutron stars or the tidal deformability from GW170817 \cite{Reed20,Liang21}. Most importantly, the lower limits of $c^2_{\rm QM}$ extracted above show clearly that the conformal limit cannot be satisfied in neutron stars.
The tension between the conformal limit and observations of neutron star has been discussed also in, e.g., Refs. \cite{Kurkela14,Bedaque15,Alsing18,Godzieba21}. In particular, the lower limit of $c^2_{\rm QM}$ was found  to be 0.55 and 0.41 in Refs. \cite{Kurkela14} and \cite{Alsing18},respectively. While Ref. \cite{Godzieba21} reported that the conformal limit must be violated at $\rho>2\rho_0$.

\begin{figure}[ht]
  \centering
   \resizebox{0.45\textwidth}{!}{
  \includegraphics{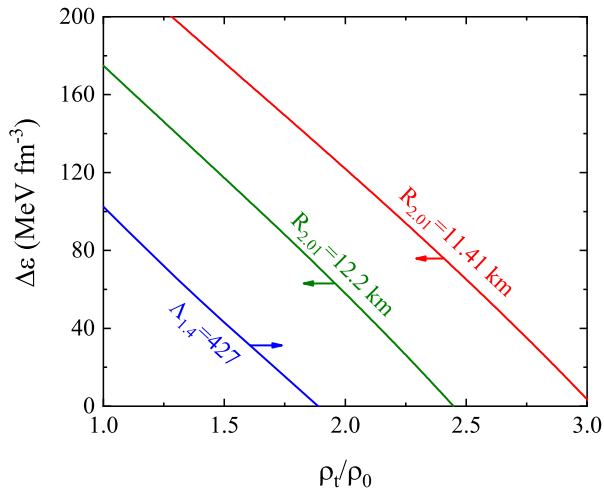}
  }
  \caption{Projections to the $\rho_t/\rho_0-\Delta\varepsilon$ plane of the intersecting line between the surfaces of $\Lambda_{1.4}=427$ and $M_{\rm max}=2.01 M_\odot$ (blue lines, labeled as $\Lambda_{1.4}=427$) and the intersecting lines between the causality condition $c^2_{\rm QM}=1$ and the surfaces of $R_{2.01}=11.41$ km (red lines, labeled as $R_{2.01}=11.41$ km) and $R_{2.01}=12.2$ km (green lines, labeled as $R_{2.01}=12.2$ km), respectively. The arrows indicate the directions that satisfying the specified observations.}\label{re}
\end{figure}

\begin{figure*}[ht]
  \centering
   \resizebox{0.9\textwidth}{!}{
  \includegraphics{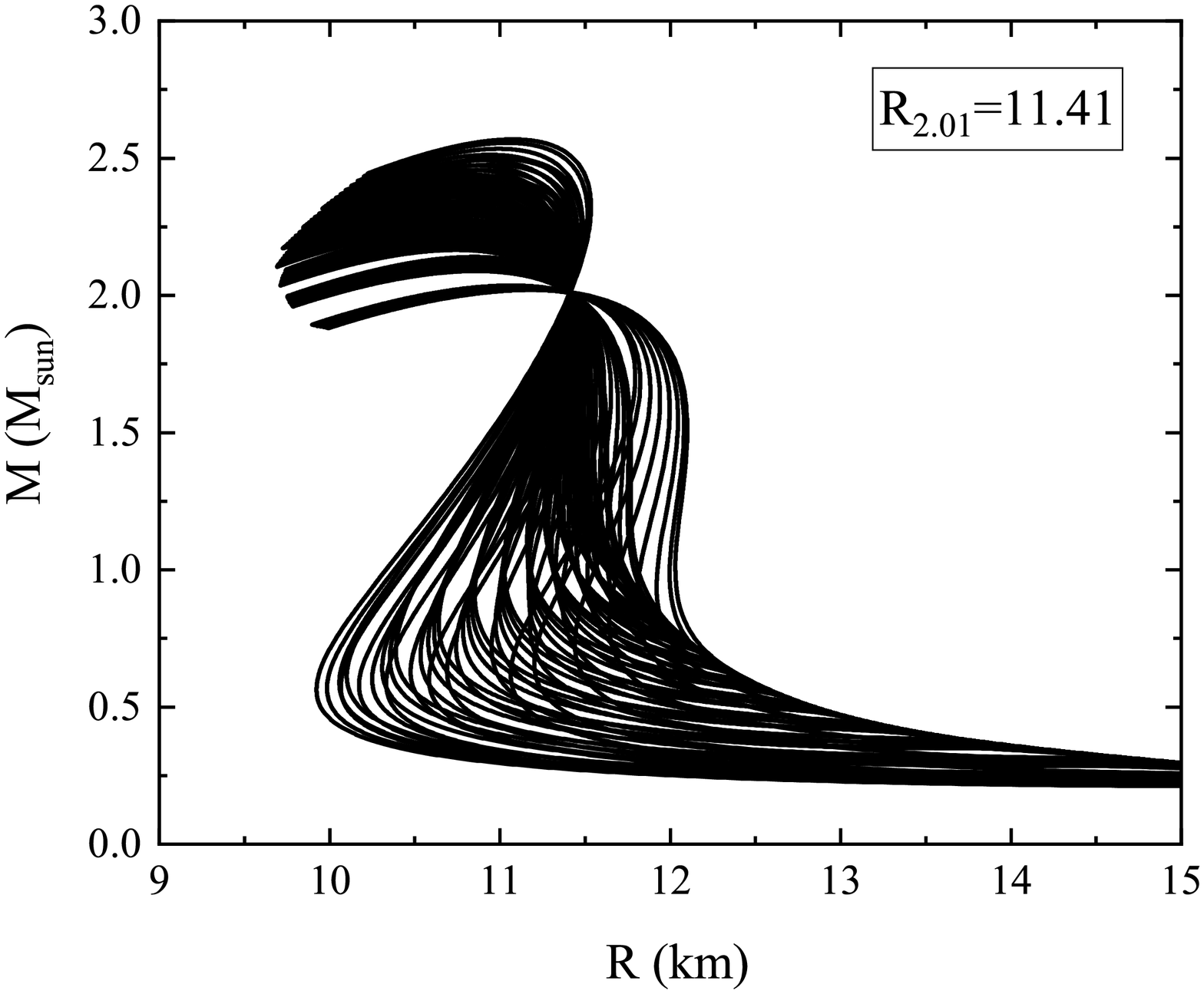}
  \includegraphics{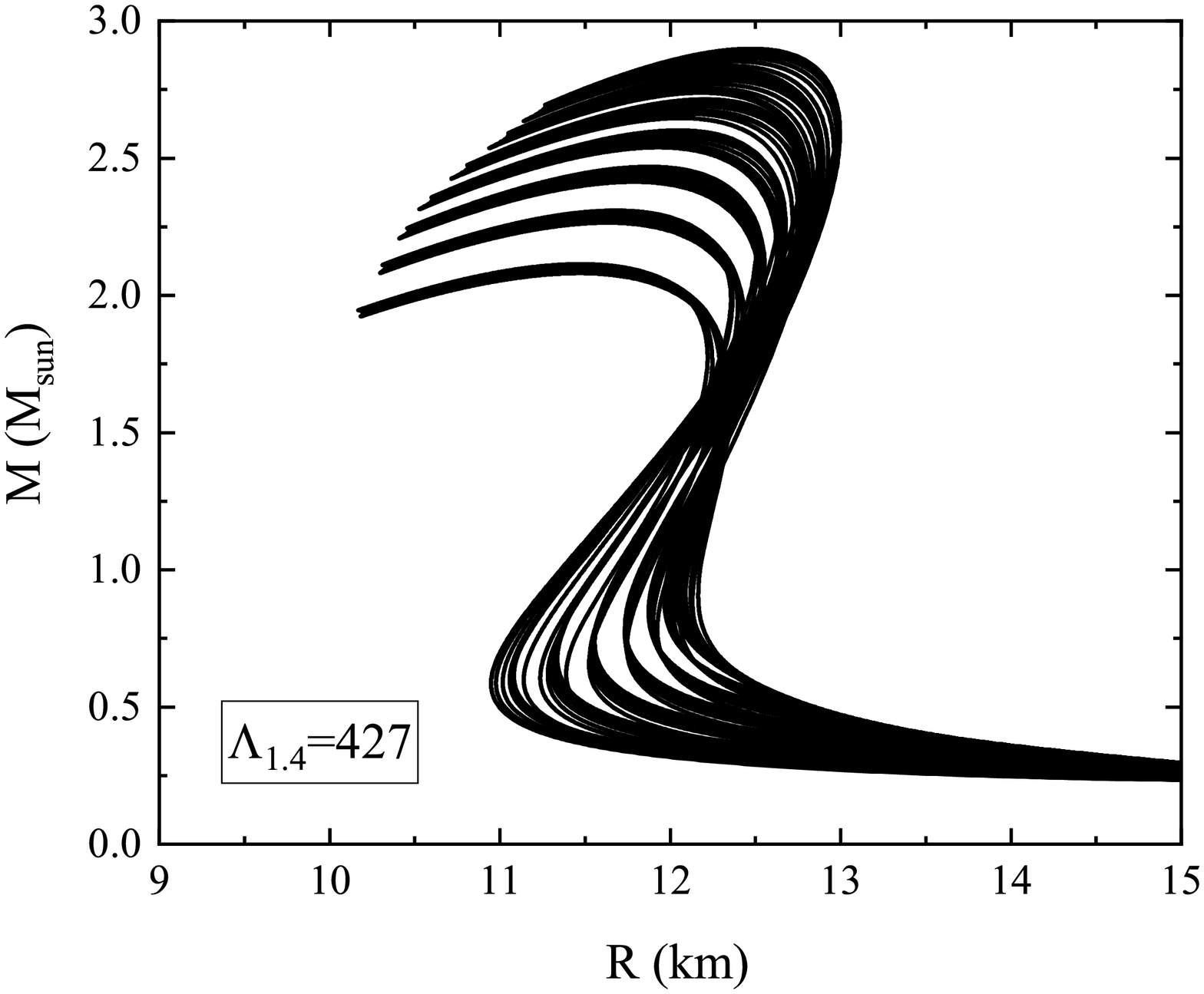}
  }
  \caption{137 mass-radius curves calculated from the parameter sets selected from the surfaces of $R_{2.01}=11.41$ km (left panel) and $\Lambda_{1.4}=427$ (right panel), respectively.}\label{twin}
\end{figure*}

Similarly, we also project the intersecting line between the surfaces of $\Lambda_{1.4}=427$ and $M_{\rm max}=2.01 M_\odot$ (blue lines, labeled as $\Lambda_{1.4}=427$), and the ones between the causality condition of $c^2_{\rm QM}=1$ and the surfaces of $R_{2.01}=11.41$ km (red lines, labeled as $R_{2.01}=11.41$ km) as well as $R_{2.01}=12.2$ km (green lines, labeled as $R_{2.01}=12.2$ km) to the $\rho_t/\rho_0-\Delta\varepsilon$ plane in Fig. \ref{re}. The arrows show the directions satisfying the indicated observations. The intersecting line between the surfaces of $\Lambda_{1.4}=427$ and $M_{\rm max}=2.01 M_\odot$ serves as the lower limit and constrains the plane from the left side, while the intersecting lines between the causality condition of $c^2_{\rm QM}=1$ and the surfaces of $R_{2.01}=12.2$ km as well as $R_{2.01}=11.41$ km provide the two upper limits from the right side. We can see that if we use the observation of $R_{2.01}=12.2$ km ($R_{2.01}=11.41$ km) as a constraint, the upper limit for $\Delta\varepsilon$ is only 175 MeV fm$^{-3}$ (231 MeV fm$^{-3}$) with the loose lower limit of $\rho_t/\rho_0=1$. Smaller values of $\Delta\varepsilon$ are favored if we use larger values for the lower limit of $\rho_t$ from Refs. \cite{Tang21,Miao20,Christian20}. This indicates that the strength of first-order hadron-quark phase transition cannot be too large. On the other hand, for $R_{2.01}=12.2$ km, the upper limit for $\rho_t$ is 2.45$\rho_0$, while it is constrained to be about 3.03$\rho_0$ with $R_{2.01}=11.41$ km. This is consistent with the finding of $\rho_t<2.5\rho_0$ in Refs. \cite{Miao20,Xie21,Liang21,Somasundaram22}.

\begin{figure}[ht]
  \centering
   \resizebox{0.45\textwidth}{!}{
  \includegraphics{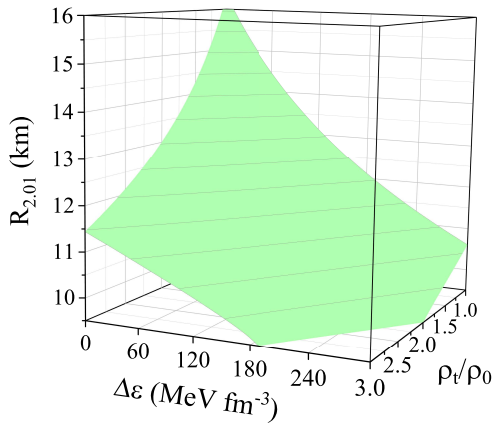}
  }
  \caption{The radius of PSR J0740+6620 versus $\rho_t/\rho_0-\Delta\varepsilon$ on the causality surface where $c^2_{\rm QM}=1$. The horizontal lines correspond to different values of $R_{2.01}$.}\label{reupperlimit}
\end{figure}

An interesting phenomenon associated with hybrid stars is the possible existence of twin stars. In this case, two stable branches with similar masses but different radii are predicted to exist for a given neutron star EOS. As shown in Fig. 3 of Ref. \cite{Alford13}, the twin star can exit for large discontinuity in energy density $\Delta \varepsilon$ and small transition pressure $p_t$. Ref. \cite{Christian20} found that $\Delta\varepsilon=350$ MeV fm$^{-3}$ is the lowest value to generate visible twin stars within a mass range larger than 0.1 M$_\odot$. In our above analysis, the upper limit of $\Delta\varepsilon$ is constrained to be 231 MeV fm$^{-3}$ for $R_{2.01}=11.41$ km. This relatively low limit on $\Delta\varepsilon$ seems to exclude the existence of twin stars. To further validate the above conjecture, we present 137 mass-radius curves calculated from the parameter sets selected from the surfaces of $R_{2.01}=11.41$ km (left panel) and $\Lambda_{1.4}=427$ (right panel) in Fig. \ref{twin}. It is clearly shown that no twin star can be observed within the parameter space constrained by the observations selected in the present work and thus the presence of twin star is disfavored. 

In addition, it is clearly shown in Fig. \ref{re} that the $\rho_t-\Delta\varepsilon$ correlation is strongly dependent on the observations while individual limits of $\rho_t$ and $\Delta\varepsilon$ cannot be constrained independently. In particular, the cross lines between $c^2_{\rm QM}=1$ and $R_{2.01}=11.41$ km (red) and $R_{2.01}=12.2$ km (green) can be well fitted by
$\Delta\varepsilon=-112.91\rho_t/\rho_0+345.86$ (MeV fm$^{-3}$) (r=0.9994)
and
$\Delta\varepsilon=-119.89\rho_t/\rho_0+296.33$ (MeV fm$^{-3}$) (r=0.9994), respectively. They have very similar slopes but different intercepts. These features indicate that on the causality surface where $c^2_{\rm QM}=1$, the $R_{2.01}$ itself may have a strong relation with $\rho_t/\rho_0$ and $\Delta\varepsilon$. To reveal this relation,  in Fig. \ref{reupperlimit} by keeping $c^2_{\rm QM}=1$, we now vary $R_{2.01}$ beyond the above two observational values from NICER. The horizontal lines correspond to constant values of $R_{2.01}$ and their projections to the bottom surface can help constrain the $\rho_t/\rho_0$-$\Delta\varepsilon$ correlation. We can see that the increase of $R_{2.01}$ can potentially constrain the $\rho_t/\rho_0$ or $\Delta\varepsilon$ more tightly. The value of $R_{2.01}$ on the causality surface can be well fitted by the following equation (with $r=0.9940$):
\begin{equation}\label{R201fit}
    R_{2.01}=7.46+12.29\rm{exp}(-\frac{\rho_t/\rho_0}{2.66}-\frac{\Delta\varepsilon}{305.40})~~(\rm{km}).
\end{equation}
This analysis here generalizes the results shown in Fig. \ref{re} and further quantifies how an accurate measurement of $R_{2.01}$ can set an upper boundary for the $\rho_t/\rho_0-\Delta\varepsilon$ correlation.

\section{Summary and Conclusions}\label{sec4}

By inverting the observational data of several neutron star observables in the 3D CSS model parameter space for quark matter while fixing all hadronic EOS parameters at their currently known most probable values, we constrained properties of the first-order hadron-quark phase transition and their correlations. This approach provides a visual representation of the effects of each model parameter on the observables or conversely the constraints on the model parameter space provided by the latter. We found that the observational constraints of $M_{\rm max}=2.01 M_\odot$, the lower radius limit of $R_{2.01}=11.41$ km and $R_{2.01}=12.2$ km are equivalent for soft quark matter EOSs (small $c_{\rm QM}^2$) and weak phase transition strengths as the radius of a neutron star with $M_{\rm max}=2.01 M_\odot$ exceeds 12.2 km in this region, and all three constraints are satisfied simultaneously. However, this convergence disappears gradually with increasing $c^2_{\rm QM}$ and $\Delta\varepsilon$.

The lower limits of $c_{\rm QM}^2$ are found to be 0.35 and 0.43 with the $R_{2.01}=11.41$ km and $R_{2.01}=12.2$ km observational constraints, respectively, which implies that the conformal limit $c_{\rm QM}^2<1/3$ cannot be satisfied in neutron stars. No correlations are observed between $c_{\rm QM}^2$ and $\rho_t/\rho_0$ or $\Delta\varepsilon$. On the other hand, the upper limit of $c_{\rm QM}^2$ is not constrained by the observations considered. The upper limits of $\rho_t/\rho_0$ and $\Delta\varepsilon$ are found to be 3.03 (2.45) and 231 (175) MeV fm$^{-3}$ with the constraints of $R_{2.01}=11.41$ km ($R_{2.01}=12.2$ km), respectively. The constraint on $\Delta\varepsilon$ indicates that the strength of first-order phase transition cannot be too large (e.g., Ref.\cite{Christian20} takes $\Delta\varepsilon>350$ MeV fm$^{-3}$ as a strong phase transition). However, the present work does not constrain the lower limits of $\Delta\varepsilon$ and $\rho_t/\rho_0$. Additionally, the $\rho_t-\Delta\varepsilon$ correlation is found to be closely dependent on the observations of $R_{2.01}$ but the individual limits of $\rho_t$ and $\Delta\varepsilon$ cannot be constrained simultaneously with the data available. Finally, considering that the upper limit of $\Delta\varepsilon$ is restricted to 231 MeV fm$^{-3}$ for $R_{2.01}=11.41$ km, the presence of twin star is deemed improbable.

Certainly, our work has limitations and caveats. Our inversion of neutron star observables is limited to the 3D CSS model parameter space. Some of the high-density hadronic EOS parameters still have large uncertainties although we used their most probable values known to us possibly with some biases. Moreover, by choice  the CSS model assumes that the hadron-quark phase transition is first order and the speed of sound in quark matter is a constant. Furthermore, possible formation of various hyperons and other particles may further complicate the situation. Finally, we found that the conformal limit will be violated in the case of a hybrid star with a first-order phase transition. However, it may still be satisfied in the case of a quark star \cite{Miao21,Traversi22} or two-families scenario \cite{Drago14,Drago16}. Nevertheless, our results obtained from using the limited observational data available indicate clearly that our approach is useful in improving our knowledge about neutron star matter. With more precise data expected to come from multi-messenger astronomy in the near future, we are hopeful that our approach will help further reveal the nature and EOS of dense neutron-rich matter.\\

\noindent{\bf Acknowledgments}\\
This work is supported in part by the U.S. Department of Energy, Office of Science, under Award Number DE-SC0013702, the CUSTIPEN (China-U.S. Theory Institute for Physics with Exotic Nuclei) under the US Department of Energy Grant No. DE-SC0009971, the National Natural Science Foundation of China under Grant No. 12005118, and the Shandong Provincial Natural Science Foundation under Grants No. ZR2020QA085.



\begin{thebibliography}{99}
\bibitem{Orsaria19} M. G. Orsaria, G. Malfatti, M. Mariani, I. F. Ranea-Sandoval, F. Garc\'{i}a, W. M. Spinella, G. A. Contrera, G. Lugones, and F. Weber, J. Phys. G: Nucl. Part. Phys. \textbf{46} 073002 (2019).
\bibitem{Weih20} L. R. Weih, M. Hanauske, and L. Rezzolla, Phys. Rev. Lett. \textbf{124}, 171103 (2020).
\bibitem{Reitze19} D. Reitze {\it et al.}, Bull. Am. Astron. Soc. \textbf{51}, 035 (2019).
\bibitem{Punturo10} M. Punturo {\it et al.}, Class. Quant. Grav. \textbf{27}, 194002 (2010).
\bibitem{Ackley20} K. Ackley {\it et al.}, Publ. Astron. Soc. Austral. \textbf{37}, e047 (2020).
\bibitem{Tang21} S. P. Tang, J. L. Jiang, W. H. Gao, Y. Z. Fan, and D. M. Wei, Phys. Rev. D \textbf{103}, 063026 (2021).
\bibitem{Yong22} G. C. Yong, B. A. Li, Z. G. Xiao, and Z. W. Lin, Phys. Rev. C \textbf{106}, 024902 (2022).
\bibitem{Liang21} A. Li, Z. Q. Miao, S. Han, and B. Zhang, Astrophys. J. \textbf{913}, 27 (2021).
\bibitem{Xie21} W. J. Xie and B. A. Li, Phys. Rev. C \textbf{103}, 035802 (2021).
\bibitem{Somasundaram22} R. Somasundaram and J. Margueron, Europhysics Letters \textbf{138}, 14002 (2022).
\bibitem{Miao20} Z. Q. Miao, A. Li, Z. Y. Zhu, and S. Han, Astrophys. J. \textbf{904}, 103 (2020).
\bibitem{Mmax} H. T. Cromartie {\it et al.}, Nature Astronomy \textbf{4}, 72 (2019).
\bibitem{Fonseca21} E. Fonseca {\it et al.}, Astrophys. J. Lett. \textbf{915}, L12 (2021).
\bibitem{Romani22} R. W. Romani, D. Kandel, A. V. Filippenko, T. G. Brink, and W. K. Zheng, Astrophys. J. Lett. \textbf{934} L17 (2022).
\bibitem{Miller21} M. C. Miller {\it et al.}, Astrophys. J. Lett. \textbf{918}, L28 (2021).
\bibitem{Riley21} T. E. Riley {\it et al.}, Astrophys. J. Lett. \textbf{918}, L27 (2021).
\bibitem{LIGO18} B. P. Abbott {\it et al.}, Phys. Rev. Lett. \textbf{121}, 161101 (2018).
\bibitem{Gorda18} T. Gorda, A. Kurkela, P. Romatschke, M. Sappi, and A. Vuorinen, Phys. Rev. Lett. \textbf{121}, 202701 (2018).
\bibitem{Kurkela10} A. Kurkela, P. Romatschke, A. Vuorinen, and B. Wu, arXiv:1006.4062.
\bibitem{Kurkela14} A. Kurkela, E. S. Fraga, J. Schaffner-Bielich, and A. Vuorinen, Astrophys. J. \textbf{789}, 127 (2014).
\bibitem{Bedaque15} P. Bedaque and A. W. Steiner, Phys. Rev. Lett. \textbf{114}, 031103 (2015).
\bibitem{Alsing18} J. Alsing, H. O. Silva, and E. Berti, Mon. Not. R. Astron. Soc. \textbf{478}, 1377 (2018).
\bibitem{Godzieba21} D. A. Godzieba, D. Radice, and S. Bernuzzi, Astrophys. J. \textbf{908}, 122 (2021).
\bibitem{Tan20} H. Tan, J. Noronha-Hostler, and N. Yunes, Phys. Rev. Lett. \textbf{125}, 261104 (2020).
\bibitem{Ecker22} C. Ecker and L. Rezzolla, arXiv:2207.04417.
\bibitem{Alford13} M. G. Alford, S. Han, and M. Prakash, Phys. Rev. D \textbf{88}, 083013 (2013).
\bibitem{Chamel13} N. Chamel, A. Fantina, J. Pearson, and S. Goriely, Astron. Astrophys. \textbf{553}, A22 (2013).
\bibitem{Zdunik13} J. Zdunik and P. Haensel, Astron. Astrophys. \textbf{551}, A61 (2013).
\bibitem{LiJJ21} J. J. Li, A. Sedrakian, and M. Alford, Phys. Rev. D \textbf{104}, L121302 (2021). Erratum: Phys. Rev. D \textbf{105}, 109901 (2022).
\bibitem{Li22} A. Li, G. C. Yong, and Y. X. Zhang, Phys. Rev. D \textbf{107}, 043005 (2023).
\bibitem{Alford15} M. G. Alford, G. F. Burgio, S. Han, G. Taranto, and D. Zappal\`{a}, Phys. Rev. D \textbf{92}, 083002 (2015).
\bibitem{Ayriyan15} A. Ayriyan, D. E. Alvarez-Castillo, D. Blaschke, H. Grigorian, and M. Sokolowski, Phys. Part. Nucl. \textbf{46}, 854 (2015).
\bibitem{Drischler22} C. Drischler, S. Han, J. M. Lattimer, M. Prakash, S. Reddy, and T. Q. Zhao, Phys. Rev. C \textbf{103}, 045808 (2021).
\bibitem{Chatziioannou20} K. Chatziioannou and S. Han, Phys. Rev. D \textbf{101}, 044019 (2020).
\bibitem{Han20} S. Han and M. Prakash, Astrophys. J. \textbf{899}, 164 (2020).
\bibitem{Agrawal10} B. Agrawal, Phys. Rev. D \textbf{81}, 023009 (2010).
\bibitem{Bonanno12} L. Bonanno and A. Sedrakian, Astron. Astrophys. \textbf{539}, A16 (2012).
\bibitem{Lastowiecki12} R. Lastowiecki, D. Blaschke, H. Grigorian, and S. Typel, Acta Phys. Pol. B Proc. Suppl. \textbf{5}, 535 (2012).
\bibitem{Kurkela10b} A. Kurkela, P. Romatschke, and A. Vuorinen, Phys. Rev. D \textbf{81}, 105021 (2010).
\bibitem{Traversi21} S. Traversi, P. Char, G. Pagliara, and A. Drago, Astron. Astrophys. \textbf{660}, A62 (2022).
\bibitem{Traversi20} S. Traversi and P. Char, Astrophys. J. \textbf{905}, 9 (2020).
\bibitem{Zhang18} N. B. Zhang, B. A. Li, and J. Xu, Astrophys. J. \textbf{859}, 90 (2018).
\bibitem{Xie19} W. J. Xie and B. A. Li, Astrophys. J., \textbf{883}, 174 (2019).
\bibitem{Xie20} W. J. Xie and B. A. Li, Astrophys. J., \textbf{899}, 4 (2020).
\bibitem{Zhang19} N. B. Zhang and B. A. Li, Euro. Phys. J. A \textbf{55}, 39 (2019).
\bibitem{Zhang19a} N. B. Zhang and B. A. Li, J. Phys. G: Nucl. Part. Phys. \textbf{46}, 014002 (2019).
\bibitem{Zhang19b} N. B. Zhang and B. A. Li, Astrophys. J. \textbf{879}, 99 (2019).
\bibitem{Zhang2020} N. B. Zhang and B. A. Li, Astrophys. J. \textbf{883}, 61 (2020).
\bibitem{Zhang2021} N. B. Zhang and B. A. Li, Astrophys. J. \textbf{921}, 111 (2021).
\bibitem{Zhang22} N. B. Zhang and B. A. Li, Euro. Phys. J. A \textbf{59}, 86 (2023).
\bibitem{Garg18} U. Garg and G. Col\`{o}, Prog. Part. Nucl. Phys. \textbf{101}, 55 (2018).
\bibitem{Shlomo06} S. Shlomo, V. M. Kolomietz, and G. Col\`{o}, Euro. Phys. J. A \textbf{30}, 23 (2006).
\bibitem{Li13} B. A. Li and X. Han, Phys. Lett. B \textbf{727}, 276 (2013).
\bibitem{Oertel17} M. Oertel, M. Hempel, T. Kl\"{a}hn, and S. Typel, Rev. Mod. Phys. \textbf{89}, 015007 (2017).
\bibitem{Li21} B. A. Li, B. J. Cai, W. J. Xie, and N. B. Zhang, Universe \textbf{7}, 182 (2021).
\bibitem{Grams22} G. Grams, R. Somasundaram, J. Margueron, and E. Khan, arXiv:2207.01884.
\bibitem{Margueron18} J. Margueron, R. H. Casali, and F. Gulminelli, Phys. Rev. C \textbf{97}, 025806 (2018).
\bibitem{Mondal17} C. Mondal, B. K. Agrawal, J. N. De, S. K. Samaddar, M. Centelles, and X. Vi\~{n}as, Phys. Rev. C \textbf{96}, 021302 (2017).
\bibitem{Somasundaram21} R. Somasundaram, C. Drischler, I. Tews, and J. Margueron, Phys. Rev. C \textbf{103}, 045803 (2021).
\bibitem{Oppenheimer39} J. Oppenheimer and G. Volkoff, Phys. Rev. \textbf{55}, 374 (1939).
\bibitem{Negele73} J. W. Negele and D. Vautherin, Nucl. Phys. A \textbf{207}, 298 (1973).
\bibitem{Baym71b} G. Baym, C. J. Pethick, and P. Sutherland, Astrophys. J. \textbf{170}, 299 (1971).
\bibitem{Reed20} B. Reed and C. J. Horowitz, Phys. Rev. C \textbf{101}, 045803 (2020).
\bibitem{Christian20} J. Christian and J. Schaffner-Bielich, Astrophys. J. Lett. \textbf{894}, L8, (2020).
\bibitem{Miao21} Z. Q. Miao, J. L. Jiang, A Li, and L. W. Chen, Astrophys. J. Lett. \textbf{917} L22 (2021).
\bibitem{Traversi22} S. Traversi, P. Char, G. Pagliara, and A. Drago, Astron. Astrophys. \textbf{660}, A62 (2022).
\bibitem{Drago14} A. Drago, A. Lavagno, and G. Pagliara, Phys. Rev. D \textbf{89}, 043014 (2014).
\bibitem{Drago16} A. Drago, A. Lavagno, G. Pagliara, and D. Pigato, Euro. Phys. J. A \textbf{52}, 40 (2016).
\end{thebibliography}
\end{document}